\documentclass[runningheads]{svmult}

\usepackage{makeidx}   
\usepackage{graphicx}  
\usepackage{subeqnar}  
\usepackage{multicol}  
\usepackage{physprbb}  
\makeindex             

\newcommand{\greeksym}[1]{{\usefont{U}{psy}{m}{n}#1}}
\newcommand{\umu}{\mbox{\greeksym{m}}}

\begin{document}
\def\aj{AJ}                   
\def\araa{ARA\&A}             
\def\apj{ApJ}                                 
\def\apjs{ApJS}               
\def\ao{Appl.Optics}          
\def\apss{Ap\&SS}             
\def\aap{A\&A}
\def\aaps{A\&AS}                
\def\aapr{A\&Ap~Rev.}                        
\def\azh{AZh}                 
\def\baas{BAAS}               
\def\jrasc{JRASC}             
\def\memras{MmRAS}            
\def\mnras{MNRAS}             
\def\pasp{PASP}               
\def\pasj{PASJ}               
\def\qjras{QJRAS}             
\def\sovast{Soviet~Ast.}      
\def\ssr{Space~Sci.Rev.}      
%
%
\title*{Globular Cluster Distances from RR Lyrae Stars  
\footnote{\footnotesize To be published in {\it Stellar Candles}, \\
Lecture Notes in Physics (http://link.springer.de/series/lnpp), \\
Copyright: Springer-Verlag, Berlin-Heidelberg-New York, 2003}}
\toctitle{Globular Cluster Distances from RR Lyrae Stars}
%
%
\titlerunning{Globular Cluster Distances from RR Lyrae Stars}
%
\author{Carla Cacciari and Gisella Clementini\inst{}}
\authorrunning{C. Cacciari \& G. Clementini}
%
%
%
\institute{INAF -- Osservatorio Astronomico, 40127 Bologna, ITALY} 
\maketitle              

\begin{abstract}

The most common methods to derive the distance to globular clusters using 
RR Lyrae variables are reviewed, with a special attention to those  
that have experienced significant improvement in the past few years. 
From the weighted average of these most recent determinations the absolute
magnitude of the RR Lyrae stars at $[Fe/H]=-1.5$ is $M_V = 0.59 \pm 0.03$,  
corresponding to a distance modulus for the LMC $(m-M)_0 = 18.48 \pm 0.05$.

\end{abstract}

\section{Introduction}

Globular clusters (GC) have traditionally been considered as good tracers
of the process that led to the formation of their host galaxy, whether 
this was a ``monolithic'' relatively rapid collapse of the primeval gas cloud, 
as described by \cite{els62}, or a ``hierarchical'' capture of smaller 
fragments on a longer time baseline, as described by \cite{sz78}. 
We refer the reader to \cite{fbh02} for a recent review on this issue. 

Therefore, knowing the distance to GCs with high accuracy is 
important in several respects: 

\begin{itemize}
\item 
Cluster distances, along with information on the dynamical, kinematic 
and chemical properties of the clusters, are essential to provide a 
complete description of the galaxy formation, early evolution and chemical
enrichment history.   

\item Accurate distances are needed in order to derive the age of GCs  
from the stellar evolution theory, i.e. by comparing the 
absolute magnitude of the Main-Sequence Turn-off (MS-TO) region in the 
Color-Magnitude diagram with the corresponding luminosity of theoretical 
isochrones. The precise knowledge of absolute ages has  important 
cosmological implications (e.g. the age of the Universe), whereas relative 
ages provide detailed information on the formation process of the host galaxy. 

\item The Luminosity Function (LF) of a GC system is one of the most powerful 
candles for extragalactic distance determinations, as it peaks at a rather 
bright luminosity ($M_V \sim -7.5$), and GCs  are numerous in both spiral and 
elliptical galaxies. 
An accurate calibration is essential both for testing the assumption 
of universality of the LF and for deriving its absolute value, hence again 
the importance of disposing of accurate distances to local GC calibrators. 
\end{itemize}

Distances to GCs can be obtained by several methods that use either the 
cluster as a whole (direct astrometry) or some 
``candle'' stellar population belonging to the cluster, e.g. Horizontal 
Branch (HB)  and RR Lyrae stars, Main Sequence stars, White Dwarfs, Eclipsing 
Binaries, the tip of the Red Giant Branch (RGB), the clump along the RGB.  
In the latter case the distance stems from the determination of 
the absolute magnitude of the selected ``candle'', which in turn may depend 
on a purely theoretical or a (semi)-empirical calibration. 

Here we review the distance determinations to GCs using only RR Lyrae 
(and HB) stars. Distance determinations to GCs using also the other possible 
methods have been reviewed by \cite{cac02}. 

\section{RR Lyrae Stars as Standard Candles} 

RR Lyrae stars have been traditionally the most widely used objects for 
the purpose of distance determination (see \cite{smi95} for a general review), 
because they are i) easy to identify thanks 
to their light variation; ii) luminous giant stars (although less bright than 
the Cepheids) hence detectable to relatively large distances; iii) typical of
old stellar systems that do not contain Population I distance indicators 
(such as the classical Cepheids), and iv) much more numerous than the 
Population II Cepheids. 
But of course the property that qualifies them as {\it standard candles} is 
their mean brightness, which has been known to be  ``nearly constant'' in 
any given globular cluster (within a narrow range of variation) 
since Bailey's work in early 1900. 

However, accurate studies have shown a few decades later that the mean 
intrinsic brightness of the RR Lyrae stars is not strictly constant: first, 
it is a (approximately linear) function of metallicity, i.e. 
$M_V(RR) = \alpha[Fe/H] +  \beta$ (\cite{san81a}; \cite{san81b}) with a 
variation of 
$\sim$0.25 mag over 1 dex variation in metallicity; 
second, even within the same cluster, i.e. at fixed metallicity, there is 
an intrinsic spread in the HB luminosity due to evolutionary effects, 
whose extent can vary from $\sim$0.1 to $\sim$0.5 mag as a function  
of metallicity (\cite{san90}); finally,  it has recently been shown that the  
luminosity-metallicity relation is not strictly linear, because it depends 
also on the HB morphology and stellar population (\cite{cap00}; \cite{dzl00}). 

Notwithstanding these aspects that introduce a significant intrinsic 
variation in the absolute magnitude of the RR Lyrae variables, these stars 
remain excellent distance indicators once these effects are properly known 
and taken into account. 

As a first approximation, and for the purpose of taking into account small 
corrections due to metallicity differences, when needed in comparing 
different results, we assume that the $M_V(RR)$-[Fe/H] relation is linear, 
with the parameters estimated by \cite{cha99} as the average of 
several methods, i.e. 
\begin{equation}
          M_V(RR) = (0.23\pm0.04)[Fe/H] +  (0.93\pm0.12) 
\end{equation}
\cite{cac02}  found the same slope and 0.92 for the zero-point.   
  
We shall now consider the main methods of absolute magnitude determination 
for RR Lyrae stars, with special attention to those that have 
experienced a substantial improvement recently.  
The aim is to estimate the most accurate and reliable value for the 
$\beta$ parameter in eq. (1), which can then 
be used to derive the distance to globular clusters and other old stellar 
systems of known metallicity. For this purpose also studies dealing with 
field RR Lyrae stars will be reviewed, on the verified assumption that 
globular cluster and field RR Lyrae stars share the same characteristics 
(\cite{cat98}; \cite{cgc00}).

\subsection{Statistical Parallaxes}

This method works by balancing two measurements of the velocity ellipsoid 
of a given stellar sample, obtained from the stellar radial velocities and 
from the proper motions plus distances, via a simultaneous solution for a 
distance scale parameter. 
The underlying assumption is that the stellar sample  can be adequately 
described by a model of stellar motions in the Galaxy.   

\cite{lay99} provided the most recent review of this method, and summarized 
the results previously obtained by various groups on 
field RR Lyraes using slightly different algorithms and assumptions but 
basically the same sample of stars and very similar input data. 
From the work of \cite{gp98}  on 147 RR Lyrae stars, which 
contains a very careful analysis and corrections for all  relevant 
biases, the average magnitude is  $M_V(RR) = 0.77\pm0.13$  at [Fe/H]=--1.6. 
A very interesting result is the analytic expression for the relative 
error in the distance scale parameter reported by \cite{pg99}  
(and references therein). 
They show that for a group of stars with a given velocity dispersion and 
bulk motion, and with observational errors smaller than the velocity 
dispersion, the relative error in the distance scale parameter is proportional 
to $N^{-1/2}$ where $N$ is the number of stars in the sample. 
For a halo stellar population such as the RR Lyraes, where observational 
errors in the radial and tangential velocities are typically 20-30 km/s 
and velocity dispersions are $\sim$ 100 km/s,  a more effective way of 
improving the results is by increasing the number of stars in the sample 
rather than improving further the quality of the velocity determinations.
\cite{gp98} tried this way by defining the radial velocity
ellipsoid using 716 metal-poor non-variable and 149 RR Lyrae stars, and 
matching it with the distance-dependent ellipsoid derived from the proper 
motions of the RR Lyraes alone. The result is $M_V(RR) = 0.80\pm0.11$  
at [Fe/H]=--1.71. 
The accuracy of this hybrid solution, however, is hardly 
any better given the bigger chance of thick disk contaminants even at  
low metallicity. We remind that \cite{bdr02}  find that the 
local fraction of metal-poor stars that might be associated with the 
Metal Weak Thick Disk (MWTD) is on the order of 30\%-40\% at 
abundances below [Fe/H]=--1.0, and a significant 
fraction  of these may extend to metallicities below [Fe/H]=--1.6.   

\cite{dr01} applied this method to a sample of 262 local  
RR Lyrae variables. They separated ``halo'' from ``thick disk'' objects  
by metallicity ($[Fe/H]<-1.0$) and kinematic criteria, and assumed an 
initial distance scale $<M_V>(RR) = 1.01 + 0.15[Fe/H]$ (i.e. $M_V$ = 0.79
at [Fe/H]=--1.5) to transform proper motions into space velocity components.
They then  determined $M_V(RR) = 0.76\pm0.12$  for the ``halo'' population at 
[Fe/H]=--1.6. This result is in agreement with the previous ones from this
method.  
However, the possibility of kinematic inhomogeneities within 
the ``halo'' sample is strongly reassessed by \cite{bor02}, 
who identify two different populations among the metal-poor subset in this 
sample of stars. 
These two spherical subsystems would have different dynamical characteristics 
and origins, the slowly rotating subsystem being associated to the Galactic 
thick disk, and the fast rotating (possibly with retrograde motion) subsystem 
belonging to the accreted outer halo. 

It is not quite clear if and to what extent the kinematic selection criteria 
adopted to separate ``halo'' and ``thick disk'' populations introduce a bias 
in the subsequent kinematic analysis of this method, and the effects on the
final result of the adopted distance scale and possible contamination from 
the MWTD stellar component. 
It is not impossible that the application of the Statistical 
Parallax method to the local RR Lyrae stars might need a more detailed 
and accurate modelling of the stellar motions in the Galaxy, as well as a 
much larger sample of stars to 
work on, in order to provide reliable and robust results.   

For the purpose of the present review we can summarize the results of the 
Statistical Parallax method as $<M_V(RR)>=0.78\pm0.12$ mag at [Fe/H]=--1.5. 
   
\subsection{Trigonometric Parallax for RR Lyr}

Trigonometric parallaxes are the most straightforward method of distance 
determination, being based on geometrical quantities independent of
reddening. Only with $Hipparcos$  trigonometric parallaxes for a good 
number of HB and RR Lyrae stars have become available. 
However, they are not accurate enough for a reliable individual distance 
determination, except for the nearest star, 
RR Lyr, for which a relatively high precision estimate of $\pi$ 
(4.38$\pm$0.59 mas) was derived by $Hipparcos$ (\cite{per97}). 
A previous ground-based estimate (i.e.  3.0$\pm$1.9 mas) was reported 
in the Yale Parallax Catalog (\cite{ypc95}). 

The new and very important result in this field is the determination 
of a more accurate parallax for RR Lyr using HST-FGS3 data 
($\pi=3.82\pm0.20$ mas) by \cite{ben02}. 
This leads to a true distance
modulus $\mu_0 = 7.09\pm0.11$ mag, or $7.06\pm0.11$ mag if one adopts instead 
the weighted average of all three parallax determinations 
$<\pi> = 3.87\pm0.19$ mas. 

RR Lyr has $<V>=7.76$ mag and [Fe/H]=--1.39 (\cite{cle95}; \cite{fer98a}). 
Depending on whether one assumes $<A_V>=0.07\pm0.03$ mag as the average 
absorption value from the 
reference stars surrounding RR Lyr, or $<A_V>=0.11\pm0.10$ mag as the 
linearly interpolated local value from the same reference stars, 
one obtains $M_V = 0.61\pm0.11$ mag or $M_V = 0.57\pm0.15$ mag, respectively.  
 
Following \cite{ben02} we adopt $M_V = 0.61\pm0.11$ mag,   
which leads to $M_V(RR)$ = 0.58$\pm$0.13 at [Fe/H]=--1.5.  This final error 
takes into account also the cosmic scatter in luminosity due to the finite 
width of the instability strip, by adding in quadrature an adopted value for 
the cosmic dispersion of 0.07 mag. This effect, which is negligible when 
many stars are involved, should be taken into account when dealing with 
individual stars.

\subsection{Trigonometric Parallaxes for HB Stars}

Since, as we discuss in Sect. 2.5, RR Lyraes are HB stars, \cite{gra98} 
adopted the approach of considering all field metal-poor HB stars 
with $Hipparcos$ values of $\pi$ in a magnitude limited sample, $V_0 < 9$. 
This selection criterium led to a sample of 22 stars, of which 10 were HB 
stars on the blue side of the instability strip, 3 were RR Lyrae stars, and 
9 were red HB stars. 
Using  the globular cluster M5 as a template to reproduce the shape of 
the HB, \cite{gra98} estimated the correction in $M_V$ to apply to each star 
in order to report it to the middle of the 
instability strip, and derived $<M_V>=0.69\pm0.10$ mag at [Fe/H]=--1.41,   
or $<M_V>=0.60\pm0.12$ mag at [Fe/H]=--1.51 excluding one red HB star 
suspected of belonging to the red giant population. 

A reanalysis of this sample was performed by \cite{pg99}, who 
eliminated all red HB stars from the sample as a prudent way to ensure that 
no contamination from the Red Giant Branch (RGB) was present, applied a
different weighting procedure by the observational errors, and considered 
the effect of intrinsic scatter in $M_V$ in the estimate of the Malmquist 
bias. Their result was $<M_V>=0.69\pm0.15$ at [Fe/H]=--1.62 
(but \cite{car00} point out that this result may be questionable 
since the metallicity scale for blue HB stars is not well determined). 
Finally, \cite{kl98}  using all stars of this sample and taking into 
account the intrinsic scatter in the HB magnitudes when correcting for the 
Lutz-Kelker effect, derived  $M_V = 0.62$ mag at [Fe/H]=--1.5. 

A final reanalysis of this problem was performed by \cite{car00}  
who provided a revised value $<M_V>=0.62\pm0.11$ at [Fe/H]=--1.5.

\subsection{Baade-Wesselink (B-W)}

This method derives the distance of a pulsating star by comparing the 
linear radius variation, that can be estimated from the radial velocity curve,
with the angular radius variation, that can be estimated from the light curve.

It is common belief that the B-W results are ``faint'', based on the large
amount of work done on field RR Lyrae stars during the past decade by several 
independent groups, and revised and summarized by \cite{fer98b}: 
\begin{equation}
M_V(RR) = (0.20\pm0.04)[Fe/H] + (0.98\pm0.05)
\end{equation}
hence $M_V(RR)$ = 0.68 at [Fe/H]=--1.5. 

This method was reapplied by \cite{cac00} to  RR Cet ([Fe/H]=--1.43, average 
value from \cite{cle95} and \cite{fer98a})
with the following improvements with respect to the previous analyses: 
\begin{itemize}

\item 
Use of various sets of model atmospheres, with and without overshooting 
treatment of convection, [$\alpha$/Fe]=+0.4;  some experimental models with 
no convection, that mimic the effects of a different treatment of convection 
e.g. the \cite{cm91} approximation, were also tried.  

\item Use of the detailed variation of gravity with phase, rather than the mean 
value; the values of log$g$ at each phase step were calculated from 
the radius percentage variation (assuming $\Delta R/<R> \sim 15\%$) plus the 
acceleration component derived from the radial velocity curve.
  
\item 
Use of new semi-empirical calibrations for bolometric corrections, based on the 
temperature scale for Population II giants defined from 
RGB and HB stars in several globular clusters using infrared colors 
(\cite{mon98}).  

\item 
Use of various assumptions on the $\gamma$-velocity, and turbulent velocity
 = 2km/s and 4km/s over all or part of the pulsation cycle.
 
\item  Use of BVRIK photometric data.  

\item 
The matching of the linear and angular radius variations was performed 
on the phase interval $0.25 \le \phi \le 0.70$ to avoid shock-perturbed 
phases.  

\end{itemize}

It was found that i) the use of K magnitudes and V--K colors provided the 
most reliable and stable results, and ii) all other options produced similar 
results within 0.03 mag, except the test case that used an unrealistically 
large amplitude of the $\gamma$-velocity curve.  

The resulting mean magnitude for RR Cet is $M_V$ = 0.57 $\pm$ 0.10 mag, 
i.e. 0.55 $\pm$ 0.12 mag when reported to [Fe/H]=--1.5, and taking into 
account the cosmic dispersion (see Sect. 2.2).

\subsection{Evolutionary models of Horizontal Branch stars}

From the evolutionary point of view, RR Lyrae stars are low-mass stars 
in the stage of core helium burning located in a well defined part of the 
HB, i.e. the temperature range approximately 5900-7400 K, known as the 
``instability strip''. Therefore theoretical models of HB stars within this 
temperature interval should in first approximation be able to describe 
the average properties of RR Lyrae variables, were they not be pulsating.   

Theoretical models (hence the HB morphology and luminosity level) depend 
significantly on assumed input parameters. The strongest dependence besides 
[Fe/H] is on the  helium abundance, but other parameters may have an effect, 
such as [CNO/Fe], peculiar surface abundances due to mixing during 
the RGB phase, diffusion or sedimentation, rotation, magnetic field strength,  
some other yet unknown factor that affects mass loss efficiency, or a 
combination of any of these, as well as theoretical assumptions such as the 
equation of state, the treatment of plasma neutrino energy loss, the correct 
treatment of conductive opacities in RGB stars, the 3-alpha reaction rate, 
etc. briefly on anything that can affect the ratio total mass vs core mass of 
the star. 

For these reasons several research groups have been actively working on 
the construction of new HB models trying to include as much improved input 
physics as possible. Without entering into the details of 
the individual choices and assumptions, for which we refer the reader to 
the original papers, we report the results found by six independent groups, 
namely \cite{dor92}, \cite{cal97}, \cite{cas99}, \cite{ferr99} (based on the 
work by \cite{scl97}), \cite{dzl00} (from outer halo globular clusters only), 
and \cite{vdb00}.  

We show in Fig. 1 how $M_V(HB)$ varies with [Fe/H], where $M_V(HB)$ is 
the mean absolute V magnitude of an HB star at log$T_{eff}$=3.85, that is 
taken  to represent the equilibrium characteristics of an RR Lyrae star 
near the middle of the instability strip.

\begin{figure}[]
\begin{center}
\includegraphics[width=.8\textwidth]{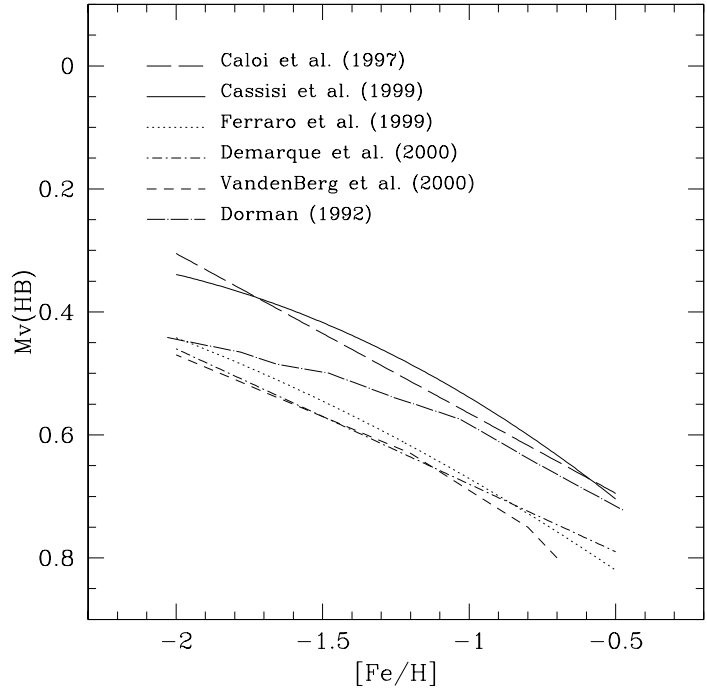}
\end{center}
\caption[]{$M_V$(HB) vs [Fe/H] from various sets of evolutionary models.}
\label{eps1}
\end{figure}

The original theoretical data usually refer to the Zero Age Horizontal Branch 
(ZAHB) rather than the average HB (or RR Lyrae) magnitude level.  
The two quantities $M_V(ZAHB)$ and $M_V(HB)$ are not identical. 
The stars in this evolutionary phase evolve rapidly away from the ZAHB: 
less than $\sim$ 10\% of their total HB lifetime is spent on the ZAHB itself, 
and the remaining time is spent at 0.1-0.2 mag brighter luminosities  
(\cite{ferr99}). 
This aspect of stellar evolution has been discussed 
in several papers (see e.g. \cite{cap87}; \cite{csj92}; \cite{cs97}).   

Therefore the real ZAHB is intrinsically poorly populated, and when a 
comparison is made with the lower envelope of the observed HBs, which would 
represent the ZAHB, this is of difficult definition because of the  
uncertainties due to small sample statistics and photometric errors. 
So the comparison between HB theoretical models and observed HBs (including 
RR Lyrae stars) is made at the (brighter) magnitude level where the stars 
spend most of their HB lifetime. This is usually taken into account by 
correcting the theoretical $M_V(ZAHB)-[Fe/H]$ relation by a fixed offset 
(of the order of 0.08-0.10 mag), or by applying an empirical correction that 
is itself a function of [Fe/H], such as the one derived by \cite{san93}:
\begin{equation}
\Delta V(ZAHB-HB)=0.05[Fe/H]+0.16 
\end{equation}  

For the sake of simplicity we apply a fixed evolutionary correction of 
--0.08 mag to $M_V(ZAHB)$, which is very close to Sandage's correction 
near the middle 
of the metallicity range at [Fe/H]=--1.5.

We can derive a few conclusions from this comparison (see Fig. 1): 

i) all models agree that the slope of the $M_V(HB)-[Fe/H]$ relation 
is not unique, i.e. this relation is {\it  not universal} and is {\it not 
strictly linear}, as  originally suggested by \cite{ccp91}.  
As a first approximation, however, all models can be roughly 
described by a linear relation with average slope $\sim$ 0.23, excluding 
the oldest set of models (\cite{dor92}) that are flatter. 

ii) As far as the zero-point is concerned, there are two families of results 
differing by $\sim$ 0.13 mag, i.e. \cite{cas99}  and \cite{cal97}  with  
$<M_V(HB)>$ =0.43 $\pm$ 0.12 at [Fe/H]=--1.5, and \cite{ferr99} (\cite{scl97}),  
\cite{dzl00} and \cite{vdb00} with $<M_V(HB)>$ =0.56 $\pm$ 0.12. Again,
\cite{dor92} models differ as they fall exactly in between these two estimates.

\subsection{Pulsation Models for RR Lyrae Stars}

\subsubsection{Visual Range}

New pulsation models have been calculated recently by \cite{cap00},   
based on non-linear convective hydrodynamical models with 
updated opacities and the classical MLT treatment of convection 
(\cite{bon97}). In combination with HB evolutionary models it is 
then possible to derive the Period-Luminosity-Metallicity relation for 
first overtone pulsators (RRc stars) at the blue edge of the instability 
strip, which in turn allows to estimate the luminosity $<M_V(RR)>$ of the 
RR Lyrae stars at the reference temperature log$T_{eff}$=3.85. 
 
The behaviour of $M_V(RR)$ vs. [Fe/H] has been found to vary with the 
HB morphology and metallicity range, and could possibly be approximated 
by a quadratic relation.  
However, for the sake of simplicity this relation can be described 
by two linear relations that, for [$\alpha$/Fe]=+0.3, are: 
\begin{equation}
M_V(RR)=(0.17\pm0.04)[Fe/H]+(0.80\pm0.10) ~at ~[Fe/H]~ < ~-1.5 
\end{equation}
and 
\begin{equation}
M_V(RR)=(0.27\pm0.06)[Fe/H]+(1.01\pm0.12) ~at ~[Fe/H] ~> ~-1.5.
\end{equation}
At the junction point [Fe/H]=--1.5 there is a discontinuity of 0.06 mag, 
the average value being $M_V(RR)$ =0.58$\pm$0.12. 

The relations expressed in eq. (4) and (5), and comparison data points 
for a number of galactic globular 
clusters, are shown in Fig. 2. Compared with the theoretical HB models shown 
in Fig.1, these pulsation models are consistent with the family of results 
that produce the fainter magnitudes. 

\begin{figure}[]
\begin{center}
\includegraphics[width=.8\textwidth]{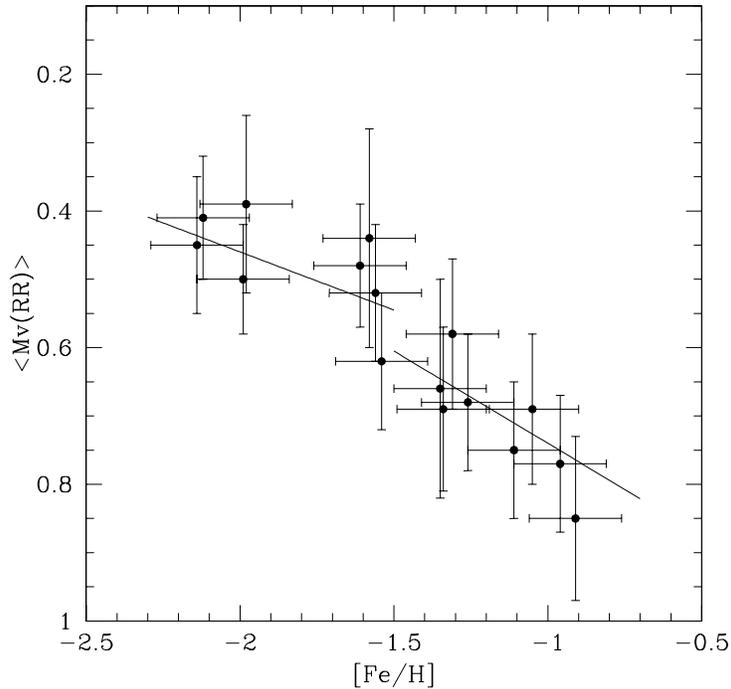}
\end{center}
\caption[]{$<M_V(RR)>$ vs [Fe/H] from the pulsation models by \cite{cap00}.
The dots represent galactic globular clusters, for comparison. }
\label{eps1}
\end{figure}

\subsubsection{Infrared Range}

Infrared (K-band) observations of RR Lyrae stars had shown already several 
years ago that there exist a relatively tight relation between period $P$ 
and mean K absolute magnitude $<M_K>$, with no (or little) dependence on 
metallicity (\cite{lon90}; \cite{lj90}; \cite{jon92}). 
These empirical relations, however, had to be calibrated on some independent 
method of absolute magnitude determination, which was usually the
Baade-Wesselink method in its various versions, hence different zero-points. 
Also the dependence on metallicity, admittedly small, was not assessed 
unambiguously. The values of $M_K$ so derived could vary on a range of 
$\sim$ 0.15 mag.  
   
Based on the same set of updated non-linear convective pulsation models 
described above, \cite{bon01}  have defined a theoretical 
Period-Luminosity-Metallicity relation in the infrared K band $(PL_KZ)$, 
which is much less sensitive to reddening and metallicity than 
the visual equivalent relation, and therefore is supposedly more accurate
(total intrinsic dispersion $\sigma_{Mk}$=0.037 mag): 
\begin{equation}
M_K=-2.071logP+0.167[Fe/H]-0.766 
\end{equation}
Note that this relation has been derived for models with solar scaled
metallicities; models with [$\alpha$/Fe]=+0.3, that are well mimicked by 
models with $[M/H]=[Fe/H]+0.21$ according to the recipe by \cite{sal93}, 
would produce fainter $M_K$ by $\sim$ 0.035 mag. 

In principle, the non-linearity of the $logL(HB)-[Fe/H]$ relation 
and the change of slope at [Fe/H]$\sim$--1.5 could be taken into account 
by defining two separate linear relations for models with [Fe/H]$<$--1.5 
and [Fe/H]$>$--1.5, respectively.  
In practice, the effect of non-linearity and the change 
of slope are negligibly small and the linear relation defined over the entire 
metallicity range that is relevant for RR Lyrae stars and globular 
clusters provides the same $M_K$ values within the errors. 

Very recently \cite{bon03}  have 
revised the above analysis and derived improved 
Period-(V--K)Color-Luminosity-Metallicity ($PCL_KZ$) 
relations for fundamental and first-overtone pulsators separately (see  
Bono, this conference, for more details). 
These new relations, also based on solar-scaled 
metallicity models, seem to give systematically brighter magnitudes by 
a few hundredths of a magnitude with respect to the $PL_KZ$ relation 
expressed in eq. (6).   

We now apply the $PL_KZ$  relation in eq. (6) to a few test cases for which 
accurate data are 
available, i.e. the RR Lyrae stars in the globular clusters M3 and 
$\omega$ Cen, the 7 field RR Lyrae stars with $[Fe/H]=-1.5\pm0.15$ for which 
the Baade-Wesselink method was applied, and the field variable RR Lyr itself. 
 
\begin{itemize}

\item{\bf M3.} \\
Seventeen non-Blazhko RRab variables and nine RRc variables in M3 have K 
photometry (\cite{lon90}). Assuming [Fe/H]=--1.47 (\cite{kra92}), 
and correcting the periods of the type-c variables to fundamental mode 
by the addition of a constant (0.127)  to their $logP$ values, we obtain a 
K distance modulus $(m-M)_K=15.03\pm0.05$, that can be considered as 
intrinsic modulus since the reddening of M3 is at most E(B--V)=0.01 mag 
(\cite{db00}), assuming  $A_V=3.1E(B-V)$ and $A_K=0.11A_V$ (\cite{ccm89}).  
For these same stars $<V>=15.64\pm0.05$ (\cite{cc01}), hence 
$<M_V(RR)>=0.58\pm0.08$.

\item{\bf $\omega$ Cen.} \\
In addition to the K photometric data obtained by \cite{lon90}  
on 30 RR Lyrae variables, new K photometry of 45 RR Lyrae variables has 
been recently obtained by \cite{ferr03}. There are no stars 
in common between these two sets of data, therefore the consistency of the 
photometric calibrations cannot be checked but {\it a posteriori}, by 
comparing the distance moduli derived  from the two sets separately. 
Metal abundances for a number of RR Lyrae variables in $\omega$ Cen have been 
recently derived by \cite{rey00}, and have been used to derive $M_K$  
via eq. (6). We obtain $(m-M)_K=13.69\pm0.09$ and $13.65\pm0.13$ 
from  \cite{lon90} and \cite{ferr03} data sets, respectively. 
The difference, well within the errors, could be ascribed to photometric 
calibration. Since \cite{ferr03} 
data are more recent and for a larger number of stars, 
we adopt their result that leads to $(m-M)_0=13.60\pm0.13$ assuming 
E(B--V)=0.13 and $A_K=0.36E(B-V)$ (\cite{ccm89}). 
This result compares very well with the 
value of $13.65\pm0.11$ obtained by \cite{tho01} using the 
eclipsing binary OGLE17 to derive the distance to $\omega$ Cen. \\
If we now consider only the variables with [Fe/H]=--1.5$\pm$0.10, they have  
$<V_0(RR)>=14.14\pm0.11$ (\cite{rey00}), hence $M_V(RR)$ =0.54$\pm$0.17 
at [Fe/H]=--1.5. 

\item{\bf Field RR Lyrae stars}\\
Approximately 30 field RR Lyrae stars  have been analyzed with the 
Baade-Wesselink method and therefore have very accurate V and K light curves 
(see \cite{fer98b} for a review). From this sample we have selected 
7 stars with $[Fe/H]=-1.5\pm0.15$, that  are listed in Table 1. 
The V and K data are from \cite{lj90} for all stars except UU Cet  
(data  from \cite{ccf92})  and WY Ant  (data from \cite{ski93}).  
Using this average value of metallicity we have then derived the 
corresponding $M_K$ values from eq. (6), the distance 
moduli and $M_V$ values. Assuming a realistic error of $\sim$ 0.1 mag on 
each individual estimate, the weighted average of these estimates is 
$<M_V>=0.61\pm0.04$ mag. 
For comparison, the average Baade-Wesselink result on these same stars, 
as reported by \cite{fer98b}, is $0.68\pm0.15$, whereas the 
application of the revised $PCL_KZ$ relation leads \cite{bon03} to 
estimate an average value of 0.54$\pm$0.03 mag.

\begin{table}
\caption{Field RR Lyrae stars with average [Fe/H]=--1.5$\pm$0.15 and good 
V and K data}
\begin{center}
\renewcommand{\arraystretch}{1.4}
\setlength\tabcolsep{5pt}
\begin{tabular}{lccccccc}
\hline\noalign{\smallskip}
Name & [Fe/H] & [Fe/H] & $<V_0>$ & $<K_0>$ & $M_K$ & $(m-M)_0$ & $M_V$ \\
     & $(1)$  & $(2)$  &         &         &       &           &       \\
\noalign{\smallskip}
\hline
\noalign{\smallskip}
 WY Ant & --1.48 & --1.32 & 10.710 & 9.621 & --0.518 & 10.139 & 0.571 \\
 RR Cet & --1.48 & --1.38 &  9.625 & 8.524 & --0.485 & 9.009 & 0.616 \\
 UU Cet & --1.28 & --1.38 & 12.005 & 10.841 & --0.565 & 11.406 & 0.599 \\
 RX Eri & --1.33 & --1.63 & 9.529 & 8.358 & --0.538 & 8.896 & 0.633  \\
 RR Leo & --1.60 & --1.37 & 10.576 & 9.660 & -0.302 & 9.962 & 0.614 \\
 TT Lyn & --1.56 & --1.64 & 9.833 & 8.630 & --0.553 & 9.183 & 0.650 \\
 TU UMa & --1.51 & --1.38 & 9.764 & 8.656 & --0.493 & 9.149 & 0.615 \\
\hline
\end{tabular}
\end{center}
$(1)$ From \cite{fer98a} \\
$(2)$ From \cite{cle95}  \\
\label{Tab1a}
\end{table}

\item{\bf RR Lyr} \\
Eq. (6) can be applied to RR Lyr, for which $[Fe/H]=-1.39\pm0.10$ 
(\cite{cle95};\cite{fer98a}), logP=--0.2466 (\cite{kaz01}),  
$<K>=6.54\pm0.04$ mag (\cite{fsb93}), and $<V>=7.76$ mag (\cite{fer98a}). 
 
The result is $M_K=-0.487$, hence a distance modulus $(m-M)_0$=7.02 or 
7.01 depending on the assumed absorption, i.e. $A_V=0.07\pm0.03$  or 
$0.11\pm0.10$ (see Sect. 2.2). 
This in turn leads to $<M_V>=0.67\pm0.11$ or $0.64\pm0.11$, respectively. 
By comparison, the results obtained by \cite{ben02}  using a 
highly accurate estimate of the trigonometric parallax are $\sim$ 0.06 mag 
brighter  (as described in more detail in Sect. 2.2).  

This same analysis, performed by \cite{bon02}  in search of the 
``pulsation parallax'' of RR Lyr, leads to $M_K=-0.541\pm0.062$ mag, whereas 
the application of the $PCL_KZ$ relation leads to  
$M_K=-0.536\pm0.04$ mag (\cite{bon03}  and this conference).  
   
If we follow \cite{ben02}  choice of reddening ($A_V=0.07$), then 
the value of $M_V$ reported to [Fe/H]=--1.5 is 0.64$\pm$0.11 mag (or 0.59 
from the $PCL_KZ$ relation). 

\end{itemize}

The weighted average of these 4 examples is $<M_V>=0.61\pm0.03$ taking the
results from the $PL_KZ$ relation in eq. (6). We have seen that the revised 
and possibly improved $PCL_KZ$ relation produces absolute magnitudes 
$\sim$0.06 mag brighter;  on the other hand all these values  would be 
fainter by $\sim$0.04 mag had non-solar-scaled metallicities 
(i.e. $[\alpha/Fe]=+0.3$) be taken into account. 

Therefore we assume as average result of this section $<M_V>=0.59\pm0.10$ 
at [Fe/H]=--1.5, where the error tries to account realistically for all 
the uncertainties still affecting this method. 

\subsubsection{Double-Mode Pulsators}

Another new result of pulsation models refers to double-mode RR Lyrae  
variables (RRd). 
From the pioneering work of \cite{vab71} on stellar pulsation 
we know that the period of a fundamental (or first overtone) pulsator is 
related with its mass, luminosity and temperature via well known formulae, 
of which we report a recent redetermination by \cite{cap98}  
that includes also some dependence on metallicity: 
\begin{equation}
logP_0 = 11.242 + 0.841logL - 0.679logM - 3.410logT_e + 0.007logZ
\end{equation}
and 
\begin{equation}
logP_1 = 10.845 + 0.809logL - 0.598logM - 3.323logT_e + 0.005logZ
\end{equation} 
The double-mode pulsators, that pulsate simultaneously in the fundamental mode 
with period $P_0$ and in the first overtone with period $P_1$, allow to 
define a relation between stellar luminosity, temperature, periods and 
metallicity, where the dependence on mass is eliminated. 
Since periods and metallicities are observed quantities and
temperatures can be derived from colors and adequate (empirical or theoretical) 
color-temperature calibrations, luminosities (hence distances) can be obtained. 
 
Based on linear nonadiabatic pulsation models and various assumptions on 
opacities and detailed element abundances, \cite{kw99} applied this method 
to the RRd stars in the Galactic globular clusters M15, M68 and IC4499. 
They did not provide direct values of $M_V$ but only a comparison with the 
results of the Fourier-decomposition method, and estimated the distance 
modulus to the LMC as 18.45-18.55.  
The same data were later reanalyzed by \cite{kov00}  along with 
$\sim$ 180 RRd variables in the LMC from the MACHO database (\cite{al00}). 
No $M_V$ values are given, but only distance moduli 
reported to the LMC. The weighted average of the four distance 
determinations to the LMC  turns out to be 
 $<(m-M)_o>(LMC)=18.50\pm0.05$. 
In this method  the main source of error is due to ambiguity in the 
zero-point $T_0$ of the color-temperature transformation. 

To derive the absolute V magnitude of RR Lyraes from the above results 
we need accurate observed V magnitudes of such stars in the LMC, with 
a good knowledge of their reddenings. 
The problem of the absolute and differential reddening 
across the LMC is a thorny problem that we cannot analyse here  
(see \cite{cle03} and this conference for a detailed discussion); here 
we have assumed the values derived by the individual authors.

A few data sets can meet these requirements, in particular:

\begin{itemize} 

\item 
\cite{cle03} report the results of observations in two fields 
of the LMC bar, where 108 RR Lyrae stars were measured. 
The data in these two fields were corrected by their respective 
reddenings, i.e. 0.086 and 0.116. 
The mean magnitude of these RR Lyrae stars at average [Fe/H]=--1.5 is  
$<V_0>=19.06\pm0.06$. Spectroscopic metal abundances were also derived, 
and the slope of the luminosity-metallicity relation was found to be 
0.214 $\pm$ 0.047, well consistent with the value of 0.23 used here. 

\item
\cite{wal92}  presented and discussed the data for 160 RR Lyraes in 6 globular 
clusters (excluding NGC 1841 that may be significantly closer to us) at average 
[Fe/H]=--1.9. The data were corrected by the respective reddening for 
each individual cluster (ranging from 0.03 to 0.13 mag with average 0.07 mag). 
The mean magnitude of these 160 RR Lyrae stars is  $<V_0>=18.98\pm0.06$. 

\item
Other data for field RR Lyrae variables in the LMC are provided by the 
MACHO experiment (\cite{al00}): 680 stars, $<V_0>=19.14\pm0.10$ at 
[Fe/H]=--1.7, assumed reddening E(B--V)=0.10.   

\item The OGLE experiment (\cite{uda00}): 6000 RR Lyrae stars, 
$<V_0>=18.91\pm0.10$ at [Fe/H]=--1.6, assumed reddening E(B--V)$\sim$0.143. \\

\end{itemize}

The error we associate to the MACHO and OGLE estimates is larger than the 
values quoted by the respective authors, but we believe it better represents 
the uncertainties due to photometric calibrations and reddening estimates 
still affecting these data sets. The large difference between these two 
results can only in part be accounted for by different values of the 
assumed reddening. Because of these uncertainties, we prefer not to use 
these results in the following considerations in spite of the very large 
number of involved stars.  

A weighted average of the first two results only, after reporting them 
to [Fe/H]=--1.5, is $<V_0>=19.07\pm0.04$. Incidentally, we note that the
average value of the last two results from the MACHO and OGLE data, that 
we have not considered because less accurate, is $<V_0>=19.06\pm0.07$ 
at [Fe/H]=--1.5, although the close agreement may be fortuitous.  

If we then use the value estimated by \cite{kov00}  
from RRd pulsators for the distance to the LMC, namely 
$<(m-M)_o>(LMC)=18.50\pm0.05$ (see also A. Walker, this conference), then 
the average magnitude of the RR Lyrae stars is $<M_V>=0.57\pm0.06$.

\subsection{Fourier Parameters of Light Curves} 

During the past decade  a series of studies were conducted, aimed at 
deriving {\em empirical} relations between the Fourier 
parameters of the light curves of RR Lyrae variables and their 
physical parameters. 
In particular, $RRc$ variables were studied by \cite{sc93a} and \cite{sc93b}, 
and $RRab$ variables were studied by Kov\'acs and collaborators in 
several papers (e.g. \cite{jur96};  \cite{kj96};  \cite{kj97}; \cite{kw01}). 

This method is based on the assumption that 
period and shape of the light curves are correlated with the intrinsic 
physical parameters of the star. There is no known theoretical justification 
for this assumption, however well defined empirical correlations do indeed 
seem to exist, and the 
quoted studies have tried to define the combinations of Fourier parameters
that best correlate with e.g. metallicity, intrinsic colors and absolute 
magnitude. The advantages of this method are potentially relevant, since its 
application  only 
requires the use of accurate V light curves, that are now becoming available 
for large numbers of variables thanks to the many photometric surveys carried 
out in the past few years for different purposes.  
In particular we consider the relation 
\begin{equation}
M_V(RR) = -1.876logP - 1.158A_1 + 0.821A_3 + K 
\end{equation}
derived by \cite{kw01}  from 383 RRab variables in globular 
clusters. This formula fits the data with $\sigma$ = 0.04 mag. The zero-point 
K, however, must be determined by some calibrator. The most recent and 
accurate estimate of K has been obtained by \cite{kin02} using RR Lyr. 
This stars is affected by the Blazhko effect (a 41-d modulation of its 
amplitude whose 
amplitude in turn varies over a 4-year period). The Fourier parameters 
of this star correspond approximately to those of a normal RRab star only 
near maximum amplitude of the primary Blazhko cycle and minimum 
amplitude of the secondary cycle (\cite{jur02}). 
By analysing data taken during one such epoch \cite{kin02} finds that 
the Fourier coefficients 
$A_1$ and $A_3$ are respectively 0.31539 and 0.09768. Using          
$M_V$=0.61 for RR Lyr (see Sect. 2.2) he then finds K=0.43. 

Based on this calibration, we apply the relation in eq. (9) to 55 normal 
RRab stars in M3  whose Fourier parameters have been recently determined 
from very accurate light curves (\cite{ccc03}).  
We find an average value $M_V=0.615\pm0.003$, with an $rms$ deviation for 
a single star of 0.02 mag. This very small {\em formal} error is purely 
statistical, and is due to the large number of stars involved in this estimate 
combined with a ``tightening'' effect by a factor $\sim$ 2 of these $M_V$ 
estimates with respect to the observed V values, whose intrinsic distribution 
has instead a $\sigma \sim$ 0.05 mag (\cite{ccc03}).  

A further test can be done using the field variable RR Cet for which  
excellent light curves are available. For this star the Fourier coefficients 
$A_1$ and $A_3$ are respectively 0.31924 and 0.10760, and logP=--0.257, 
hence $M_V$=0.63 mag to which we can associate an $rms$ error of 0.05 mag. 

Both M3 and RR Cet have very similar metallicity, $[Fe/H]=-1.47$ and 
$-1.43$ respectively,  and if we report the average of these two 
determinations  to $[Fe/H]=-1.5$ we obtain $M_V=0.61\pm0.05$ mag.

\section{Summary and Conclusions}  

We have reviewed the methods of absolute magnitude determination for RR Lyrae
variables, that can be used for distance determinations to globular clusters 
and all other stellar systems containing this type of stars. 

We have adopted $M_V(RR)$ at [Fe/H]=--1.5 as the most convenient reference 
parameter (i.e. zero-point magnitude) for distance determination, 
assuming in first approximation that the dependence of $M_V(RR)$  on 
metallicity [Fe/H] is linear with a slope $\sim$ 0.23. 

We collect in the following Table 2 all the determinations of $M_V(RR)$ 
described in the previous sections.   
If we take the weighted average of these  results, we obtain 
{\bf $<$M$_V$(RR)$>$ = 0.59$\pm$0.03} mag (r.m.s. error of the mean) at 
{\bf [Fe/H]=--1.5}.



\begin{table}
\caption{Summary of $M_V(RR)$ determinations at [Fe/H]=--1.5 from the methods 
described in the text.}
\begin{center}
\renewcommand{\arraystretch}{1.4}
\setlength\tabcolsep{5pt}
\begin{tabular}{lcl}
\hline\noalign{\smallskip}
Method & $M_V(RR)$ & Reference \\
       & at [Fe/H]=--1.5 & \\
\noalign{\smallskip}
\hline
\noalign{\smallskip}
Statistical parallaxes & 0.78$\pm$0.12 & Sect. 2.1  \\
Trigonometric parallaxes (RR Lyr) & 0.58$\pm$0.13 & Sect. 2.2 \\
Trigonometric parallaxes (HB stars) & 0.62$\pm$0.11 & Sect. 2.3 \\
Baade-Wesselink (RR Cet) & 0.55$\pm$0.12 & Sect. 2.4  \\
HB stars: evolutionary models - bright & 0.43$\pm$0.12 & Sect. 2.5 \\
HB stars: evolutionary models - faint & 0.56$\pm$0.12 & Sect. 2.5  \\
Pulsation models (visual) & 0.58$\pm$0.12 & Sect. 2.6  \\
Pulsation models ($PL_KZ$) & 0.59$\pm$0.10 & Sect. 2.6\\
Pulsation models (RRd) & 0.57$\pm$0.06 & Sect. 2.6  \\
Fourier parameters & 0.61$\pm$0.05 & Sect. 2.7 \\
\hline
Weighted average value & 0.59$\pm$0.03 & \\
\hline
\end{tabular}
\end{center}
\label{Tab1a}
\end{table}

The last two values of the list, from the double-mode 
pulsators and Fourier parameters, have smaller errors than the other results 
mainly because of the large number of stars considered by these two methods. 
If we do not wish to attach to them more weight than they probably 
deserve for intrinsic merits, and consider instead a typical error of 
$\pm$0.10 mag for each of them, the previous average result and related 
error remain unchanged. Similarly, we may want to consider the results from 
the HB evolutionary models separately for the bright and faint groups: 
this would make a difference of at most 0.01 mag on the weighted average. 

We note that the average value derived above is virtually identical to 
the value obtained by \cite{cha99}, only the error is now significantly 
smaller.  Also \cite{cac02}  obtained a very similar average result 
(0.57 $\pm$ 0.04) by including the values from other distance determination 
methods, e.g. Eclipsing Binaries and Main-Sequence fitting to local Sub-Dwarfs. 
We might be tempted to conclude that we are approaching a robust result on 
this issue. \\  

Using the value of $<V_0>=19.07\pm0.04$ at [Fe/H]=--1.5 estimated in 
Sect. 2.6 
for the RR Lyrae variables in the LMC,  this translates into a distance 
modulus to the LMC $(m-M)_0$ = 18.48$\pm$0.05. 
We refer to  A. Walker and G. Clementini (this conference) for independent 
distance determinations to the LMC. \\

It may be interesting to compare the present result with two other $M_V(RR)$ 
or $M_V(HB)$ determinations that are important for different reasons: \\

i) The method of globular cluster Main-Sequence fitting to local Sub-Dwarfs 
(SD) is considered probably the most accurate and reliable method presently 
available, provided adequate precautions are taken in analyzing the data. 
The most recent results are given by \cite{gra02},  who have 
reanalyzed three clusters (47 Tuc, NGC6397 and NGC6752) using the most 
accurate data and assumptions, in particular 
high resolution (VLT-UVES) abundances and accurate photometry and reddening 
for MS and SD stars, all in the same scale and with the same treatment. 
The result obtained by \cite{gra02}  is $M_V(RR)$ = 0.61$\pm$0.07 mag
at [Fe/H]=--1.5. \\

ii) Color-Magnitude diagrams have been derived for several globular 
clusters in M31 using $HST$ data (\cite{ric01}).  
From these an estimate of the  mean HB 
magnitudes at the middle of the instability strip could be derived. 
These estimates are of course affected by significantly 
larger errors than any of those discussed in this review, however they are 
important because they allow to compare the same type of results in the 
Milky Way and in M31, in the framework of the similarities and differences 
between these two galaxies. \\ 
A preliminary analysis of 17 clusters shows that a slope $\sim$0.23 is 
adequate to describe the $V_0(HB)-[Fe/H]$ relation, and 
$<V_0(HB)>=25.06\pm0.15$ at [Fe/H]=--1.5. The corresponding value of 
$M_V(HB)$ depends on the assumed distance to M31: if we assume the 
widely used value $(m-M)_0=24.43\pm0.06$ by \cite{fm90}, 
based on the Cepheid distance scale, then $M_V(HB)$ = 0.63$\pm$0.16 mag. 
An  independent  distance determination to the centroid of the M31 globular 
cluster system by \cite{hol98}, by fitting theoretical isochrones to the 
observed red giant branches of 14 globular clusters in M31, yields  
$(m-M)_0=24.47\pm0.07$, hence $M_V(HB)$ = 0.59$\pm$0.17 at [Fe/H]=--1.5.  \\
 
It is reassuring to see that these results, in spite of the different 
intrinsic accuracy and statistical weight, agree with the average value 
estimated from the data listed in Table 2 within $1\sigma$.

\bigskip

{\bf Acknowledgements: } We are very grateful to T. Kinman for providing 
his calibration based on RR Lyr in advance of publication. We thank A. Gould 
for reminding us of cosmic scatter when dealing with individual stars.

\end{document}